\documentclass[twocolumn,prd,groupedaddress, amsmath, amsfonts,fleqn, showpacs]{revtex4}

\usepackage{graphicx}
\usepackage{graphics}
\usepackage{qsymbols}
\usepackage{array}
\usepackage{subeqnarray}
\usepackage{bm}

\def\refe@jnl#1{{#1}}
\def\aj{\refe@jnl{Astron.~J.}}                  
\def\araa{\refe@jnl{Annu.~Rev.~Astron.~Astrophys.}}
\def\apj{\refe@jnl{Astrophys.~J.}}                 
\def\apjl{\refe@jnl{Astrophys.~J.~Lett.}}          
\def\aap{\refe@jnl{Astron.~Astrophys.}}            
\def\mnras{\refe@jnl{Mon.~Not.~R.~Astron.~Soc.}}   
\def\prd{\refe@jnl{Phys.~Rev.~D}}        
\def\fcp{\refe@jnl{Fund.~Cos.~Phys.}}  
\def\physrep{\refe@jnl{Phys.~Rep.}}   
\def\physlett{\refe@jnl{Phys.~Lett.}}

\newcommand{\beq}{\begin{equation}}
\newcommand{\eeq}{\end{equation}}
\newcommand{\beqa}{\begin{eqnarray}}
\newcommand{\eeqa}{\end{eqnarray}}
\newcommand{\beqs}{\begin{subeqnarray}}
\newcommand{\eeqs}{\end{subeqnarray}}
\newcommand{\nn}{\nonumber}
\newcommand{\pd}{\partial}

\renewcommand{\vec}{\bm}

\begin{document}
\title{Cosmological Magnetogenesis driven by Radiation Pressure}
        
 \author{Mathieu Langer} 
 \author{Jean-Loup Puget}
 \author{ Nabila Aghanim}

 \affiliation{Institut d'Astrophysique Spatiale\\ Universit\'e Paris-Sud\\ F-91405 Orsay cedex\\ France\\}
\date{}
\vspace{0.5truecm}

\begin{abstract}
The origin of large scale cosmological magnetic fields remains a mystery, despite the continuous efforts devoted to that problem. We present a new model of magnetic field generation, based on local charge separation provided by an anisotropic and inhomogeneous radiation pressure. In the cosmological context, the processes we explore take place at the epoch of the reionisation of the Universe. Under simple assumptions, we obtain results (i) in terms of the order of magnitude of the field generated at large scales and (ii) in terms of its  power spectrum. The amplitudes obtained ($B\sim 8.~10^{-6} ~\mu{\rm G}$) are considerably higher than those obtained in usual magnetogenesis models and
 provide suitable seeds for amplification by adiabatic collapse and/or dynamo during structure formation. 

\end{abstract}

\pacs{98.80.-k; 98.65.-r; 95.30.Qd}

   \maketitle
%

\section{Introduction}
According to observations, magnetic fields are pervading the whole Universe. In particular, both synchrotron emission and Faraday rotation measurements reveal a magnetic field present on large scales. On galactic scales \cite{beck}, on galaxy cluster scales \cite{eo} and even on larger scales \cite{kronberg}, the observed magnetic field strengths range from $0.1$ micro-Gauss to a few tens of micro-Gauss.

The origin of the magnetic field is, however, still a mystery. Clearly, the fields cannot be generated directly with the observed amplitudes since, in that case, they would strongly affect the formation of structures \cite{rr}.
Therefore, the fields must arise in the form of weak seeds which are subsequently, in some way, strongly amplified after the onset of structure formation. Various mechanisms have been proposed for the generation of such seeds. Basically, they can be sorted in two types.

In cosmology, the mechanisms of the first type operate before matter-radiation decoupling. They take advantage of high energy physics processes such as inflation \cite{ratra,dimo,maroto} or phase transitions in the early Universe \cite{sigl,grario}. However, these mechanisms are still unsatisfactory in two respects. First, they do not provide a real prediction since the generated fields can be as weak as $10^{-65}$ Gauss and as strong as $10^{-9}$ Gauss. Moreover, a property shared by fields created before decoupling is that the stronger their amplitude, the smaller the scale on which they are generated 
(see Battaner \& Lesch~\cite{bl}, and references therein). Furthermore, under the assumption of \textit{frozen magnetic flux}, these fields decay following cosmological expansion as $(1+z)^2$. Therefore, fields of primordial origins (generated before nucleosynthesis, say) are extremely weak on galactic scales and are very strong on much smaller scales. Finally, extremely tight constraints on pre-BBN magnetic fields from gravity wave production have been recently derived \cite{chiara}, possibly ruling out a majority of the proposed models.

The mechanisms of the second type produce magnetic fields after matter-radiation decoupling. They rely essentially on the battery effect \cite{battery} which was first introduced to explain stellar magnetic fields and is basically a consequence of charge separation. Several authors explored the same mechanism in various cosmological contexts \cite{lazarian,lc,khanna}.
The intensity of the fields generated is found to be roughly of the order of $10^{-20}$ Gauss. These weak seeds are then amplified on galactic scales by a dynamo sustained by turbulence and differential rotation in protogalaxies. As noticed by Grasso \& Rubinstein \cite{GR}, this type of mechanism can hardly be responsible for magnetic fields on galaxy cluster scales. Consequently, we would have to deal with different mechanisms to explain the origin of fields in galaxies and clusters, which exhibit similar properties. Moreover, this mechanism cannot really account for the strong fields observed in objects at high redshifts where galaxies did not have enough time to rotate and amplify the seeds efficiently.

Finally, whatever the problems of each type of mechanisms, the seeds produced are generally small. As such, they require the action of a strong process capable of amplifying the magnetic amplitudes up to the values observed in galaxies.  The dynamo mechanism evoked above has long been considered suitable for that purpose but it is, however, subject to controversy  nowadays.
In fact, it has recently been pointed out \cite{ketal} that dynamo  tends to amplify the small scale magnetic fields in a more efficient way than fields on  large scales. The consequence of this is that small scale fields reach rapidly dynamically important strengths and stop, in back-reaction, the dynamo mechanism itself.
This happens long before large scale fields can reach the observed equipartition values.

 Significant progress has been achieved very recently and the back-reaction of small scale fields is now better understood. In particular, the importance of magnetic helicity conservation in dynamo quenching has been demonstrated in ideal cases \cite{fb02,blackman,branden}. However, for their application to real astrophysical systems such as galaxies, these theories require further developments. Small scale fields thus remain a possible threat to dynamo amplification mechanisms.

In this paper, we propose a mechanism which appears to be relatively free of the problems outlined above. Applied to cosmology, it operates after decoupling in the natural context of inhomogeneous reionisation, for a redshift $z$ between 6 and 7 \cite{becker, go}. Similarly to the battery effect, it relies upon local charge separation provided in our case by radiation pressure. The physics involved is fairly simple and was nevertheless left aside until now, surprisingly enough.
Notice however that Subramanian \textit{et. al.}~\cite{subra} proposed  a mechanism (which has been explored numerically by Gnedin \textit{et. al.}~\cite{gnedin}) for the generation  of magnetic field seeds at reionisation, and obtained magnetic fields of the order of $10^{-20} - 10^{-19}$ Gauss on galactic scales. Their solution for magnetogenesis relies explicitly on the actual battery effect which requires non-parallel temperature and electron density gradients. While exploring the physics at a comparable epoch, we consider here a  different and more efficient process, as will appear below.

 In the following section, we describe in details the mechanism of our model. Then, in Sect. 3, we present the results obtained for the cosmological case, emphasising on the magnetic field power spectrum and its order of magnitude at protogalactic and galactic scales.
We also apply our mechanism to the case of  interstellar magnetic fields.
Finally, we discuss the model and its limitations in Sect. 4.

\section{Physical processes}
\begin{figure}[t]
\includegraphics[width=80mm]{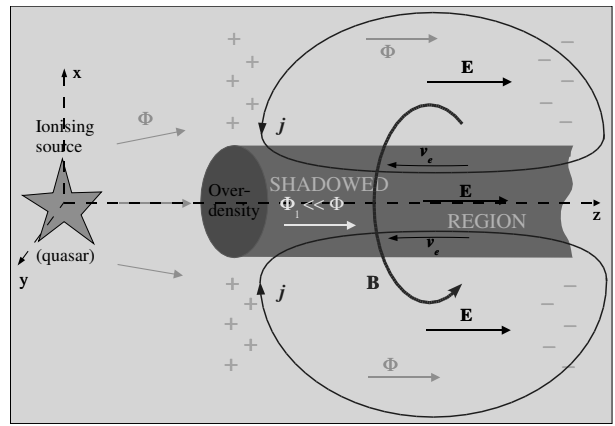}
\caption{\label{schema} Schematic view of the magnetic field generation process. The source emits a radiation flux $\vec{\Phi}$ which, at some distance, defines a preferred direction $(Oz)$ (anisotropy). It ionises the diffuse gas, and the radiation pressure locally separates electrons from ions. A static uniform electric field $\vec{E}$ is generated in response to $\vec{\Phi}$ and the electrostatic force balances the radiation pressure. The situation is different behind an overdensity: In the shadowed region, the radiation flux is weaker ($\vec{\Phi}_1 \ll \vec{\Phi}$), there is no static equilibrium and free electrons are set into motion ($\vec{v}_\text{e}$). The electric current $\vec{j}$ created this way closes up outside of the shadowed tube and induces a magnetic field $\vec{B}$ on the scale of the overdensity.}
\end{figure}

\subsection{Description}\label{desc}
Consider a radiation flux propagating through an ionised plasma. Far away enough from the dominant source,  the radiation is anisotropic and its propagation direction defines a local preferred axis. 
Since the scattering cross sections of electrons and ions are very different ($\sigma_\text{T} \propto \rm{mass}^{-2}$), the radiation pressure combined to the anisotropy of the flux lead to a charge separation, which in turn creates uniform electric fields on large scales compensating the difference in radiation pressure.

 Furthermore, the radiation propagates through a medium where the matter is distributed inhomogeneously and some of the over-densities act as screens for the radiation. Moreover, the sources of the radiation are not distributed homogeneously. Thus, because of the source distribution and the gas density inhomogeneities, the radiation flux and particularly its ionising part, is anisotropic and inhomogeneous far away from the sources. To first order, the inhomogeneities of the flux remain unchanged along the propagation axis over regions which correspond to typical distances between matter density fluctuations. In such regions, the variations in the flux allow electrons to move and electric currents  to develop. The latter finally induce a magnetic field on the length scales over which the radiation flux varies (see figure \ref{schema} for an illustrative description of the processes as they would work for one single source and one single density inhomogeneity).

The situation described above is likely to
exist at the time of the reionisation of the Universe when the very first sources (population III stars or first galaxies and quasars) turn on and ionise the neutral primordial gas, preferentially in the low density regions. Collapsing or collapsed structures play the r\^ole of the matter inhomogeneities leaving their imprints in the radiation flux.

\subsection{Formalism}

 We study the situation in a region far enough from the dominant ionising source where the direction of the ionising flux $\vec{\Phi}$ can be considered constant. This direction defines the axis $(Oz)$ (see fig. \ref{schema}). The anisotropy of the flux is essential in that it leads to a large scale charge separation which in turn generates a large scale  electric field. We could have equivalently considered a globally isotropic flux with an additional anisotropic component. 

Further, in our model, we describe the anisotropic ionising flux $\vec{\Phi}$ as a homogeneous background $\vec{\Phi_0}$ with  an inhomogeneous part superimposed on it, 
\beq
\vec{\Phi} = \vec{\Phi}_0 + \vec{\Phi}_1
\eeq
where $\vec{\Phi}_1$ accounts for the inhomogeneities which are taken to be small with respect to the constant part $\vec{\Phi}_0 = \langle \vec{\Phi} \rangle$, namely  $\vec{\Phi}_1 = f(\vec{r})~\vec{\Phi}_0$ with $f(\vec{r}) \ll 1$. As we wrote above, the region which we study is far enough from the dominant source. At such a distance, the density fluctuations in the medium along the line of propagation $(Oz)$ of the radiation have left imprints on the ionising flux which is therefore inhomogeneous. In other words, the $\vec{\Phi}_1$ part of the flux takes into account the inhomogeneities due to both the distribution of sources and the variations of the flux along the line of propagation $(Oz)$ due to the variations in the matter density. In the regions of study, the properties of the medium are supposed not to vary significantly along the line of propagation. Provided this assumption, the spatial dependency of $\vec{\Phi}_1$ is confined to the plane $(xOy)$, which implies that derivatives with respect to $z$ can be taken equal to zero, and $f(\vec{r})= f(x,y)$. Finally, we also consider the stationary regime, and  therefore all the time derivatives are nil as well.

\subsection{Maxwell equations}
Together with the assumptions specified above, the usual Maxwell equations
\beqs
&\vec{\nabla}\cdot \vec{E}&  = 4\pi\rho  \slabel{c}\\
&\vec{\nabla} \times \vec{E}& = -\frac{1}{c}\pd_t \vec{B} \slabel{f} \\
&\vec{\nabla} \cdot \vec{B}&  = 0  \slabel{g} \\
&\vec{\nabla} \times \vec{B}& = \frac{4\pi}{c}\vec{j} + \frac{1}{c}\pd_t\vec{E}\slabel{a}
\eeqs
($\rho$ and $\vec{j}$ are the charge and the current densities respectively, $c$  being the speed of light) reduce to the following simple relations:
\beqa
\pd_x E_x +\pd_y E_y = 4\pi \rho
\eeqa
 for the Coulomb equation (\ref{c}), 
\beqs
&\pd_x E_y& - \pd_y E_x = 0 \\
&\pd_y E_z& = 0 {\rm, }~\pd_x E_z = 0
\eeqs
 for the Faraday equation (\ref{f}), 
\beqa
\pd_x B_x + \pd_y B_y = 0
\eeqa
 for the Gauss equation (\ref{g}) and
\beqs\label{a2}
\pd_xB_y &-&\pd_y B_x = \frac{4\pi}{c} j_z\\
\pd_yB_z &=& \frac{4\pi}{c} j_x  {\rm, }~ -\pd_x B_z =\frac{4\pi}{c} j_y
\eeqs
 for the Amp\`ere equation (\ref{a}).
Relations (\ref{a2}) indicate that the electric currents constitute the only sources of magnetic field in the problem. We consider that the medium is globally neutral. The charge balance between free electrons, of number density $n_\text{e}$ and charge $q_\text{e}$, and ions, of density $n_\text{i}$ and charge $q_\text{i}$, implies $n_\text{i}q_\text{i}+n_\text{e}q_\text{e} = 0$. The electric current is $\vec{j} = n_\text{e} q_\text{e} \vec{v}_\text{e} + n_\text{i} q_\text{i} \vec{v}_\text{i} = n_\text{e} q_\text{e} (\vec{v}_\text{e} - \vec{v}_\text{i})$. In the following, we suppose, as customary in similar problems, that the ion velocity is negligible with respect to the electron velocity, and we write therefore $\vec{j} = n_\text{e} q_\text{e} \vec{v}_\text{e}$.

Combining the Coulomb and the Amp\`ere relations with the $\pd_t = 0$, $\pd_z = 0$ conditions, the charge conservation equation takes the following form:
\beq
\pd_x j_x +\pd_y j_y = 0.
\eeq
Within the stationary assumption, if we take the curl of equation (\ref{a}) we obtain 
\beq\label{lapl}
\vec{\nabla}^2\vec{B} = - \frac{4\pi}{c}\vec{\nabla} \times \vec{j}
\eeq
which, because of our assumptions on the spatial dependency of $\vec{\Phi}_1$, reduces to
the following relations between the magnetic field and the current density:
\beqs\label{useful}
&\vec{\nabla}^2 B_x &= -\frac{4\pi}{c} \pd_y j_z \\
&\vec{\nabla}^2 B_y &=  \frac{4\pi}{c} \pd_x j_z \\
&\vec{\nabla}^2 B_z &=  \frac{4\pi}{c} (\pd_y j_x - \pd_x j_y)
\eeqs
We will later combine the latter relations with the equation of motion.

\subsection{Equation of motion}
Given the relation outlined in the previous section between the current $\vec{j}$ and the electron velocity $\vec{v}_\text{e}$, the equation of motion for the electrons can be re-written in terms of the current evolution:

\beq\label{motion}
\frac{m_\text{e}}{n_\text{e} q_\text{e}} \frac{{\rm d}\vec{j}}{{\rm d}t} = q_\text{e}\vec{E} + \frac{1}{n_\text{e}  c} \vec{j} \times \vec{B} - \frac{m_\text{e}\nu_\text{c}}{n_\text{e} q_\text{e}} \vec{j} + \vec{F}_\text{ext}
\eeq
where $m_\text{e}$ is the electron mass, $\nu_\text{c}$ is the mean collision frequency associated with momentum transfer between electrons and ions, and $\vec{F}_\text{ext}$ stands for the external forces. The derivative on the left hand side of this equation is the whole Lagrangian derivative, i.e. $\frac{{\rm d}}{{\rm d}t} = \pd_t + \vec{v} \cdot \vec{\nabla}$ which, in the stationary regime, reduces to $ \frac{{\rm d}}{{\rm d}t} = \vec{v} \cdot \vec{\nabla}$. On the right hand side of equation (\ref{motion}), the first two terms are the Lorentz force, while the third term represents the Ohmic current dissipation which depends on the physical state of the medium. For a partially ionised gas, we should add to (\ref{motion}) a term representing the force exerted by the neutrals on the electrons. However, we concentrate on the case where the medium is ionised enough for the collisions with the neutrals to be negligible. Thus, we do not consider here this effect. Neglecting also gravity, the only external force applied on the electrons is the radiation pressure force, which in terms of the ionising flux reads $\vec{F}_\text{ext} = \frac{h\nu}{c}~\sigma_\text{T}~\vec{\Phi}$ ($h\nu$ is the mean energy of ionising photons and $\sigma_\text{T}$ is the Thomson scattering cross section).  

\subsection{Linearised problem}
Assuming the inhomogeneities $f(x,y)$ of the ionising flux to be small, we can linearise equation (\ref{motion}). Identifying the different terms of the same order leads to the following relations:
\beqs
E_x &=& \frac{m_\text{e}\nu_\text{c}}{n_\text{e}q_\text{e}^2} ~j_x\slabel{elec1}\\
E_y &=& \frac{m_\text{e}\nu_\text{c}}{n_\text{e}q_\text{e}^2} ~j_y\slabel{elec2}\\
E_z &=& - q_\text{e}^{-1}~\frac{h\nu}{c}~\sigma_\text{T}~\Phi_0 \slabel{force}
\eeqs
and 
\beq
j_z(x,y) = \frac{n_\text{e} q_\text{e}}{m_\text{e} \nu_c}~\frac{h\nu}{c}~\sigma_T~\Phi_0~f(x,y)\label{current}
\eeq
where second order terms have been neglected.

Equation (\ref{force}) shows simply the equilibrium between the radiation pressure and the electrostatic force exerted on the free electrons, whereas equation (\ref{current}) relates the induced current to the flux inhomogeneities. The current changes direction depending on the sign of the deviations $f(x,y)$ from the mean value of the flux, and therefore the current lines are closed over scales similar to those of the flux variations in the $(xOy)$ plane and comparable to the coherent scale of the anisotropy along $z$.

Taking the curl of equation (\ref{motion}) and keeping again only the first order terms gives
\beq
\vec{\nabla}^2\vec{B} =  -\frac{4\pi}{c} ~\frac{h\nu}{c} ~\sigma_\text{T} ~\vec{\nabla}\times\left(\frac{n_\text{e} q_\text{e}}{m_\text{e} \nu_\text{c}}~\vec{\Phi}\right)
\eeq
where equation (\ref{lapl}) has been used to eliminate the current.
Again, due to the particular geometry of our problem and taking profit of equations (\ref{elec1}) and (\ref{elec2}), the equation above can be rewritten as
\beqs\label{magnet}
\vec{\nabla}^2 B_x &=& - \pd_y ~\left[\kappa f(x,y)\right] \slabel{magx} \\\
\vec{\nabla}^2 B_y &=& ~~\pd_x ~\left[\kappa f(x,y)\right]  \slabel{magy} \\
\vec{\nabla}^2 B_z &=& ~~0 \slabel{magz}
\eeqs
with $\kappa = \frac{4\pi}{c} ~\frac{n_\text{e} q_\text{e}}{m_\text{e} \nu_\text{c}}  ~\frac{h\nu}{c}  ~\sigma_\text{T} ~\Phi_0$. The generated magnetic field thus depends on the physical state of the medium through the friction coefficient $\frac{n_\text{e} q_\text{e}}{m_\text{e} \nu_\text{c}}$. It depends also on the flux through both its mean intensity $\Phi_0$ and its spatial fluctuations $f(x,y)$.

\section{Results}
 The solutions to (\ref{magnet}) give the generated magnetic field. Formally, we obtain 
\beqs
B_x(\vec{r}) &=& - \iint \ln{|\vec{r}' - \vec{r}|} ~\pd_y ~(\kappa f(\vec{r})) ~{\rm d}^2\vec{r}' \\
B_y(\vec{r}) &=&  ~\iint \ln{|\vec{r}' - \vec{r}|} ~\pd_x ~(\kappa f(\vec{r})) ~{\rm d}^2\vec{r}'
\eeqs
Of course, it is also possible to find a solution for $B_z$. Integrating equation (\ref{magz}) would require precise boundary conditions. But in the model presented here, there are \textit{a priori} no such obvious physical conditions, and imposing for instance $B_z = 0$ at some distance would introduce an arbitrary non physical scale. Actually, recall that the study we present here applies only  to the domain where an axis $(Oz)$ is defined by the anisotropic flux and in which all the $z$-dependencies can be projected onto the $(xOy)$ plane. In reality, many zones with  locally constant $\vec{\Phi}$ may coexist and the local directions of the flux would be independent of each other. At the common borders of these regions, the three components of the field would get mixed. In that case, the only pertinent boundary conditions for $B_z$ would be fixed by the $x$ and $y$ components of the fields generated in similar surrounding regions. Given the restrictions provided by our initial assumptions, including such a complex description of the situation is obviously beyond the scope of this paper.

Therefore, our intention here is not to find the exact value of the magnetic field at each point in space but rather to address the question of the origin of large scale magnetic fields. For that purpose, provided the relations found in (\ref{magnet}), it is more interesting to derive global properties of the magnetic field such as its order of magnitude and its power spectrum.

\subsection{Order of magnitude}\label{ofm}
In the previous section, we found the equations which relate the transverse components of the magnetic field to the ionising flux perturbations (Eqs. \ref{magnet}) . From those relations, we can estimate the order of magnitude of the generated field. Relations (\ref{magnet}) can be expressed as 
\beq\label{om0}
B \approx \kappa ~f~R   
\eeq
where $R$ is the characteristic scale over which $f$ (hence $B$) varies. 

As we wrote earlier, the friction coefficient $\frac{n_\text{e} q_\text{e}}{m_\text{e} \nu_\text{c}} = \frac{\sigma_\text{e}}{q_\text{e}}$, appearing in $\kappa$, depends on the physical state of the plasma. In particular, it varies according to the amplitude of the magnetic field, and two regimes can be distinguished. When the electron-ion collision frequency $\nu_\text{ei}$ is greater than the cyclotron frequency $\omega_\text{c}$, we are in the non-saturated regime, $\nu_\text{c} = \nu_\text{ei}$ and the electrical conductivity is given by $\sigma_\text{e} = \sigma_0 \approx 6.5 ~10^6 ~T^{3/2}$ e.s.u \cite{lang} for a fully ionised medium. But as soon as the magnetic field is strong enough to have $\omega_\text{c} \gg \nu_\text{ei}$ (i.e. in the saturated regime), the actual electron-ion collision frequency, thus also the conductivity, are modified anisotropically. When the particle radius of gyration becomes smaller than the mean free path, the situation is unaffected along the magnetic field lines whereas collisions transverse to the lines are less likely to occur. Therefore, to the first order, the conductivity along the magnetic field lines is unchanged, whereas in the transverse direction (along the axis $(Oz)$ in our case), $\sigma_\text{e}$ takes the form $\sigma_\text{e} \sim \sigma_0 \frac{\nu_\text{ei}^2}{\omega_\text{c}^2+\nu_\text{ei}^2}$ (known as the Pederson conductivity).

 Finally, inserting these expressions into (\ref{om0}) and rewriting $h\nu \Phi_0$ as $\frac{L}{4\pi D^2}$ ($L$ is the luminosity of the ionising source and $D$ the distance to it), we end up with the following order of magnitude estimates:
\beqa\label{fi_ns}
B &\approx& 3.09~10^{-5} ~f ~\left(\frac{T}{10^{4}~{\rm K}}\right)^{3/2}\nn \\
&&\times \frac{R}{100~{\rm kpc}} ~\frac{LD^{-2}}{10^{-8}~\rm{W.m^{-2}}}~~\rm{ Gauss}
\eeqa
for the magnetic field seed generated in a fully ionised plasma in the non-saturated regime, and
\beqa\label{fi_s}
B &\approx& 3.14~10^{-2} ~f^{1/3} ~\left(\frac{T}{10^{4}~{\rm K}}\right)^{1/2}\nn \\
&&\times \left[\frac{R}{100~{\rm kpc}} ~\frac{LD^{-2}}{10^{-8} ~\rm{W.m^{-2}}} \right]^{1/3} B_\text{s}^{2/3}~~\rm{ Gauss}
\eeqa
in the saturated regime ($B_\text{s}=2\pi\nu_\text{ei} m_\text{e} c e^{-1}$ is the saturation magnetic field which depends, through $\nu_\text{ei}$, on the density and the temperature of the medium).

\subsection{Application to cosmological magnetic fields}\label{cosmo}
We now want to apply the result obtained above to the case of cosmological magnetic fields. Since the plasma which we consider is (almost)  fully ionised at $z\sim 7$, we take the temperature $T$ to be of the order of $10^4$ K. For a flat cosmology in a matter dominated Universe with $\Omega_\text{b} h_0^2 = 0.020$ \cite{burles}, the gas density is $n(z) = 3.8 ~10^{-7}~(1+z)^3~\rm{cm^{-3}}$ and the corresponding saturation magnetic field is $B_\text{s} \sim 1.4 ~10^{-11} ~T^{-3/2}~(1+z)^3$ Gauss which, for a temperature of $10^4$ K, gives $B_\text{s} \sim 1.4 ~10^{-17} ~(1+z)^3$ Gauss.

The term $LD^{-2}$ 
can be estimated by calculating the number of photons necessary to reionise the Universe. Basically, at the reionisation redshift $z_\text{ion}$, the integrated  photon density production must be equal to the gas density:
\beq\label{ionis}
\int_0^{t(z_\text{ion})} \mathcal{L} N_\text{s}(z) ~dt = n(z_\text{ion})
\eeq
where $\mathcal{L}$ is the number density of photons emitted per unit time and $N_\text{s}(z)$ is the density of emitting sources per logarithmic interval of luminosity at redshift $z$. The major contribution to reionisation comes from photons emitted roughly at redshift $z_\text{ion}$, so that the above relation  can be rewritten as
\beq
\mathcal{L} N_\text{s}(z_\text{ion}) t_\text{u}(z_\text{ion}) = n(z_\text{ion}).
\eeq
In a flat, matter dominated cosmology, the age of the Universe is $t_\text{u}(z)= \frac{2}{3}H_0^{-1} ~(1+z)^{-3/2}$. Therefore, 
\beq
\mathcal{L}N_\text{s}(z_\text{ion}) = 5.7 ~10^{-7} H_0 (1+z_\text{ion})^{9/2}.
\eeq
The term $LD^{-2}$ is then written
\beqa
LD^{-2} &=& 8.1 ~10^{-11}  h_0^{2/3}\left(\frac{h\nu}{20~{\rm eV}}\right)^{2/3} \nn \\
& & \times \left(\frac{L}{10^{12}L_{\odot}}\right)^{1/3} (1+z_\text{ion})^{3} ~{\rm W.m^{-2}} \label{ld}
\eeqa
where $h_0 = \frac{H_0}{100 ~\rm{kpc.Mpc^{-1}.s^{-1}}}$ is the reduced Hubble constant and $h\nu$ is the photon energy (in units of 20 eV). Taking a luminous quasar as the nearest ionising source (responsible for the anisotropy of the flux), $L \approx 10^{12} L_\odot$. With $h\nu \approx 20$ eV, and considering $h_0 \approx 0.70$, equation (\ref{ld}) gives
\beq
\frac{LD^{-2}}{10^{-8} {\rm W.m^{-2}}} \approx 6.38 ~10^{-3}~(1+z_\text{ion})^3
\eeq
At a redshift $z_\text{ion} \sim 7$, this corresponds to  $D\approx 3.12$ Mpc. Notice that this is an underestimate of the mean distance to the nearest quasar since we assumed implicitly in equation (\ref{ionis}) that luminous quasars were alone responsible for the ionisation of the medium.

With the values of the density and the temperature reached at high redshifts, it is easy to check that the estimated non-saturated magnetic field is greater than the saturation value $B_\text{s}$.
 Therefore, to obtain the correct order of magnitude for $B$, we have to consider the saturated regime.
Inserting the figures derived above into equation (\ref{fi_s}), we obtain 
\beq\label{estim}
B \sim 1.56 ~10^{-14}~\left(\frac{R}{100~{\rm kpc}}\right)^{1/3} ~(1+z_\text{ion})^3 ~{\rm Gauss}
\eeq
where we have taken $T\approx 10^4$ K and the degree of fluctuations in the ionising flux $f \sim 10 \%$.

The length $R$ represents the typical scale of the inhomogeneities in the radiation flux. As we already mentioned, the flux inhomogeneities take into account simultaneously both the density fluctuations in the medium and the variations due to the spatial distribution of ionising sources. 
Therefore, the scale $R$ in our calculations is not easy to define precisely without proper statistical estimates of the source and gas density distributions.  Nevertheless, we can at least estimate reasonable bounds for it. On large scales, the mechanism proposed here remains valid as long as the flux holds its property of anisotropy. Clearly, on scales larger than the mean distance $D$ between the considered ionising sources, this is no more the case, and therefore $D$ defines approximately an upper limit for $R$. On small scales, the variations of the flux would be due essentially to the matter density fluctuations. More precisely, the lower limit for $R$ would be set by the typical size of the smallest halos which could survive photoevaporation by the ionising flux itself.
However, for the simple order of magnitude evaluation intended here, it is sufficient to choose $R$ as the size of a protogalaxy, $R\approx 100 ~{\rm kpc}$, in which case, the magnetic field seed at protogalactic scales is
\beq\label{seed}
B \sim 8. ~10^{-12} ~{\rm Gauss.}
\eeq

This result is very interesting since it is several orders of magnitude higher than  fields found in usual plasma physics magnetogenesis models.

One may ask whether turbulence will affect the magnetic field generation mechanism. Indeed, in the vicinity of the ionising source, the gas motion is highly turbulent, but in fact, the actual effect of turbulence is not easily predicted. Presumably, turbulent motions will strongly fold the field lines on small scales leading  to local magnetic field amplification and/or to magnetic reconnection. Moreover, turbulence may even prevent electric currents from settling over large distances, thus inhibiting the mechanism presented here.

 However, even in presence of turbulence, the ionising radiation will be able to escape the vicinity of the source. Thus, far enough from the source, which is the case of interest explored here, electric currents will be able develop and induce magnetic fields.
 On large scales, the growth of the density fluctuation field is in the linear regime and the collapse of a large scale structure is still in its early stages where the fluid motion is not turbulent and relative displacements are small. Nevertheless, given the relatively large amplitude derived in Eq. \ref{estim}, we might want to check that  the magnetic pressure will not affect the collapse of the structure. 
The velocity $\delta\vec{v}_k$ of gas within a density perturbation $\delta_k$ on a scale $R=2\pi/k$ is given by 
\beq
\left|\delta\vec{v}_{k} \right| \simeq \frac{H}{k}\delta_{k} \simeq 11.1\, \delta_k (1+z)^{3/2}\frac{R}{1\,\text{Mpc}}\,\text{km.s}^{-1}
\eeq
in the linear regime and for a flat, matter dominated Universe with $H_0\sim 70\,\text{km.s}^{-1}\text{.Mpc}^{-1}$. Thus, the kinetic energy density is written
\beqs
E_\text{K} &=& \frac{1}{2}\rho_\text{b}\delta_k\left|\delta\vec{v}_{k} \right|^2 \\
E_\text{K} &\approx& 2.3\,10^{-19}\delta_k^3 \left(\frac{R}{1\,\text{Mpc}}\right)^2\nn\\
&&\times(1+z)^6\,\text{erg.cm}^{-3},
\eeqs
where $\rho_\text{b} \approx 3.76\,10^{-31}(1+z)^3\,\text{g.cm}^{-3}$ is the mean baryon density, whereas the magnetic energy density is 
\beqs
E_\text{B} &=& \frac{B^2}{8\pi} \\
E_\text{B} &\approx& 2.1\,10^{-28}f^{2/3}\left(\frac{R}{1\,\text{Mpc}}\right)^{2/3}\nn\\
&&\times(1+z)^6 \,\text{erg.cm}^{-3},
\eeqs
from Eq. \ref{estim}.
The ratio of the two energy densities,
\beq
\frac{E_\text{B}}{E_\text{K}} \approx 9.1\,10^{-10}f^{2/3}\delta_k^{-3}\left(\frac{R}{1\,\text{Mpc}}\right)^{-4/3},
\eeq
does not depend explicitely on redshift. At a scale $R\sim 1\,\text{Mpc}$ and considering $f\sim \delta_k \sim 0.1$, this gives $E_\text{B}/E_\text{K} \approx 2.\,10^{-7}$. According to the last relation above, $E_\text{B}$ and $E_\text{K}$ become comparable on scales smaller than a parsec. At such small scales, structures are deep in the non-linear regime which we do not consider here.
Therefore, the pressure provided by the magnetic field generated by our mechanism on large scales is well below the kinetic energy density. This guarantees that the collapse of the structure will not be modified by the presence of the magnetic field. Thus, as stated above, as long as the structure is in the linear regime, the collapse will not affect our mechanism.
 
On the contrary, the collapsing gas will drag the field lines and amplify the field strength eventually. In fact, the amplification by adiabatic collapse gives a galactic magnetic field sufficiently high to serve as seed for dynamo. The galactic value obtained from (\ref{seed}) after adiabatic collapse (and before further contraction by radiative cooling) is 
\beq\label{galactic}
B_\text{gal} = \delta_\text{c}^{2/3} B ~\sim~ 2.7 ~10^{-10}~{\rm Gauss}
\eeq
(where $\delta_\text{c}\sim 200$) from which dynamo amplification needs less time than in usual models in order to reach the fields of a few micro-Gauss observed in galaxies.

\subsection{Magnetic field power spectrum}\label{power}
Let us start from the magnetic field two-dimensional spatial correlation function
\beq
\xi(\vec{r}) = \langle \vec{B}(\vec{r})\cdot\vec{B}(\vec{r} + \vec{r}')\rangle
\eeq
where $\vec{r}$ and $\vec{r}'$ are confined to the $(xOy)$ plane. Inserting the Fourier transform of $\vec{B}$ in that expression, we end up with
\beqa
\xi(\vec{r}) &=& \left( \frac{L}{2\pi} \right)^2 \int \vec{B}_{\vec{k}}\cdot\vec{B}_{-\vec{k}} e^{-i\vec{k}\cdot\vec{r}} {\rm d}^2\vec{k} \nn\\
\xi(\vec{r}) &=&  \left( \frac{L}{2\pi} \right)^2 \int \left| \vec{B}_{\vec{k}} \right|^2  e^{-i\vec{k}\cdot\vec{r}} {\rm d}^2\vec{k} 
\eeqa
where the last line comes from the reality of $\vec{B}$. The quantity we are now interested in is the power spectrum of the magnetic field $P_\text{B}(k)$, \textit{i.e.} the angular average of the spectral correlation function $\tilde{\xi}(\vec{k})$:
\beq
P_\text{B}(k) = \frac{1}{2\pi} \int_0^{2\pi}\tilde{\xi}(\vec{k}) {\rm d}\theta 
\eeq
where $\tilde{\xi}(\vec{k})$ is the Fourier transform of $\xi(\vec{r})$. It is straightforward to see that $\tilde{\xi}(\vec{k}) = \left| \vec{B}_{\vec{k}} \right|^2$ so that the power spectrum is written
\beq
P_\text{B}(k) = \frac{1}{2\pi} \int_0^{2\pi} \left| \vec{B}_{\vec{k}} \right|^2 {\rm d}\theta.
\eeq
Restricting ourselves to the case of the non-saturated regime, we can consider  the conductivity to be constant and we can replace $\sigma_\text{e}$ by $\sigma_0$ in $\kappa$.
If we then Fourier transform equations (\ref{magx}) and (\ref{magy}), we get
\beqs
k^2 ~B^x_k &=& ~\kappa ~k_y ~f_k \\
k^2 ~B^y_k &=& - \kappa ~k_x ~f_k 
\eeqs
which finally allows us to obtain
\beq
P_\text{B}(k) = \kappa^2 \frac{\left| f_k \right|^2}{k^2} \label{ps}
\eeq

Thus, the mechanism we propose generates magnetic fields preferably on large scales whereas small scale fields are partially suppressed. This is very interesting with respect to dynamo theories. Since in our model the small scale fields are partially damped, the problem of their amplification evoked in the introduction is less crucial. However, the exact expression of the magnetic field power spectrum $P_\text{B}(k)$ depends on the shape of the flux power spectrum $\left| f_k \right|^2$.

The fluctuations $f(x, y)$ of the ionising flux are directly related to the fluctuations of the optical depth according to 
\beqa
f(x,y) &=& e^{-\tau_\text{l}(x,y)} -1 \\
& {\rm where }& \tau_\text{l}(x,y) = -\sigma_\text{i} \bar{n}_\text{h} \int^l_0 \delta(\vec{r}) dz
\eeqa
($\sigma_\text{i}$ is the ionisation cross-section, $\bar{n}_\text{h}$ is the mean particle density and $\delta$ represents the density fluctuations) and the fluctuation power spectrum can be expressed in terms of the power spectrum of  the optical depth $\left| \tau_k \right|^2$.

Following Bartelmann \& Schneider~\cite{wl} in their derivation of the Limber's formula in Fourier space, we relate the latter to the density power spectrum $\left| \delta_k \right|^2$ as
\beqa
\left| \tau_k \right|^2 &\propto& \int \frac{{\rm d}^2\vec{k}_\perp'}{2\pi}\delta_\text{D}(\vec{k}_\perp' - \vec{k}_\perp) \nn \\
&&\times \int {\rm d}k_\parallel' \left| \delta_{k'} \right|^2 \left( \frac{2}{k_\parallel' l} \sin{\frac{k_\parallel' l}{2}} \right)^2 .
\eeqa
The last integral in the equation above represents in Fourier space  the projection on the $(xOy)$  plane of the matter density over a distance $l$. Density fluctuations on longitudinal scales larger than $l$ do not contribute to the projected power spectrum and the integration  therefore must be carried over the interval $ k_\parallel \in \left[\frac{2\pi}{l}, +\infty \right[$. 

\begin{figure}[t]
\rotatebox{90}{
\includegraphics[height=86mm]{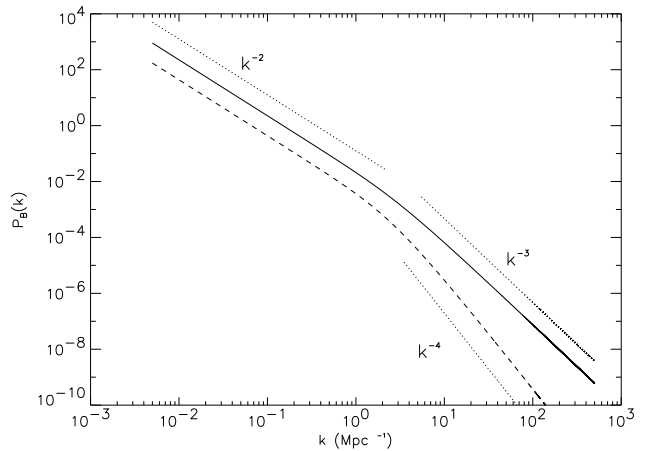}
}
\caption{\label{spectre} Power spectrum of the generated magnetic field in the non-saturated regime. A power law $\left| \delta_k \right|^2 \propto k^n$ power spectrum has been assumed for the density fluctuations. Solid line: $n=-1$; dashed line: $n=-2$. Normalisation is arbitrary. Slopes are indicated by dotted lines.}
\end{figure}

In the optically thin limit, $P_\text{B}(k)\propto \left|\tau_k \right|^2 /k^2$. Assuming a power law density spectrum $\left| \delta_k \right|^2 \propto k^n$, we calculated the magnetic field power spectrum numerically in the non-saturated case. Results shown on figure \ref{spectre} for the cases $n=-1$ and $n=-2$ assume the integration depth $l=1$ Mpc. The contribution of density fluctuations on scales larger than $l$ to the optical depth fluctuations in the plane $(xOy)$ is independent of the scale. Therefore $\left| f_k \right|^2$ is scale independent and $P_\text{B}(k) \propto k^{-2}$ for $k < 1/l$. Note however that this part of the power spectrum is of little pertinence since it corresponds to  scales larger than $l \sim 1$ Mpc. These scales are comparable to, or larger than the mean distance between ionising sources and at those scales, our assumptions break down (lost flux anisotropy). On scales smaller than $l$, the magnetic field power spectrum is strongly decreasing. Depending on the spectral index $n$ chosen for the density power spectrum, small scale magnetic fields are strongly damped. For $n\approx -1$, corresponding roughly to the spectral index on galaxy cluster scales, the magnetic field power spectrum goes like $k^{-3}$, whereas on galactic scales where $n\approx -2$, small scale magnetic fields are damped even more strongly,  $P_\text{B}(k) \propto k^{-4}$. This result makes our magnetogenesis model a very interesting ingredient for dynamo amplification theories for which small scale fields seem to be problematic, as explained in the introduction.

We have nevertheless to bear in mind that the results above are derived in the non-saturated regime. A similar study is currently undertaken for the saturated case. However, we might expect that even in the saturated case, small scale fields will be largely damped.

\subsection{Application to the Interstellar Medium}
 Magnetic fields in galaxies are organised on large scale and have intensities of a few microgauss (see, e.g., Beck \cite{beck1}, and references therein). 
Observations show that the structure of the magnetic field in the Interstellar Medium (ISM) 
is correlated with the matter distribution comforting the growing evidence that the ISM is structured not only by gravity but also by supersonic alfv\'enic turbulence.

The situation presented in Sect. \ref{desc} occurs naturally in the ISM, especially for clouds in the vicinity of OB star associations. It is thus legitimate to ask whether our mechanism can contribute significantly to the observed magnetic field strengths and possibly account for the poloidal geometry of magnetic fields observed around molecular filaments \cite{heiles}, as suggested by figure \ref{schema}.

As long as the electron-ion collisions dominate over electron-neutral collisions, we can apply directly  the results derived in Sect. \ref{ofm}. 
The electron-neutral collision cross section is of the order of a few
$10^{-15} {\rm cm}^2$ \cite{kt}.
In the neutral diffuse ISM where $T \sim 1. ~10^4$ K, $n_\text{H} \sim 1. ~{\rm cm}^{-3}$ and
$\frac{n_\text{e}}{n_\text{H}} \sim 1. ~10^{-2}$, the electron-neutral and the electron-ion collision frequencies are $\nu_\text{en} \approx 1.5~10^{-9} ~{\rm s}^{-1}$ and 
 $\nu_\text{ei} \approx 9.6~10^{-7} ~{\rm s}^{-1}$ respectively. Therefore, the ionised component dominates in the diffuse neutral ISM.
In interstellar clouds, where hydrogen is neutral but carbon is still ionised, $T = 30$ to $100$
K, $n_\text{H} = 100$ to $1000 ~{\rm cm}^{-3}$ and $\frac{n_\text{e}}{n_\text{H}} \sim 1. ~10^{-4}$, we have
$\nu_\text{en} = 1.~10^{-8}$ to $9.~10^{-8}~{\rm s}^{-1} $ while $\nu_\text{ei} = 7.~10^{-4}$ to $3.~10^{-2} ~ {\rm s}^{-1}$ and the contribution of ions is even more dominant. Thus, we can ignore the presence of neutrals and apply equations (\ref{fi_ns}) and (\ref{fi_s}).

When the electron-ion collisions dominate, the unsaturated magnetic
field is independent of the density and depends only on the temperature.
Taking the photon flux given by a $10^{6}~ L_{\odot}$ OB association at 10 pc
and a transverse gradient created by a 1 pc interstellar cloud, we get
\beq
B_\text{unsat} = 1.25 ~10^{-10} ~f~ T^{3/2} ~\text{Gauss}
\eeq
or $ B_\text{unsat} \approx  1.25 ~10^{-4} ~f$ Gauss in  the diffuse medium and
$ B_\text{unsat} \approx 1.5 ~10^{-8} ~f$ Gauss in clouds.
The critical fields $B_\text{crit}$ where the Pederson conductivity saturates are
$2.~10^{-10}$ to $1.~ 10^{-8}$  and $3.5~ 10^{-13} $ Gauss in the
clouds and the diffuse medium respectively.
We thus see that the saturation plays little r\^ole in  clouds but
reduces by a large factor the field in the diffuse medium.
The saturated magnetic fields which are produced by our mechanism are
\beq 
 B_\text{sat}  =  (B_\text{unsat}  B_\text{crit}^{2})^{1/3}
\eeq
that is $B_\text{sat}= 2.4 ~10^{-10} ~f^{1/3}$ Gauss in the diffuse medium and 
$2. ~10^{-9} ~f^{1/3}$ to $1.5~10^{-8}~f^{1/3}$ Gauss in clouds. For the specific case of molecular clouds, the induced magnetic field is roughly one order of magnitude smaller than in clouds.
We therefore conclude that in the context of the ISM, our mechanism 
is a subdominant source of magnetic fields as compared with galactic dynamo and turbulence.  In fact, the regime in the ISM is highly non linear, and the amplitude of relative motions of particles is large. Therefore, the processes discussed in Sect. \ref{cosmo} dominate, and turbulence will possibly not allow electric currents to establish on long distances. The proposed mechanism fails therefore to explain the apparent poloidal structure of fields around filaments.

\section{Conclusions}
We presented a new model for the generation of large scale magnetic fields. The mechanism proposed relies upon the simple physics of local charge separation which occurs naturally in the cosmological context of reionisation, at a redshift $z \sim 7$. As we have seen, the first luminous objects, such as quasars, emit a radiation which ionises the gas filling the Universe. The anisotropy of the ionising flux leads to the generation of electrical fields. The distribution of the ionising sources together with the gas density inhomogeneities imply that the ionising flux is inhomogeneous. The fluctuations in the flux generate electric currents which in turn induce a magnetic field on the scale of the fluctuations. 

With only a couple of simple initial assumptions, we could derive interesting properties for the generated magnetic fields. 
In particular,
the fields produced in our model have amplitudes several orders of magnitude higher than in usual plasma physics magnetogenesis models. These fields are therefore suitable to serve as seeds on large scales for adiabatic collapse amplification up to $10^{-10}$ Gauss in galaxies. Such large values represent an improvement for further amplification by dynamo mechanisms since less time is  required in order to account for magnetic fields observed in high redshift galaxies.

Furthermore, relating the magnetic field power spectrum to the matter density power spectrum, we have shown that magnetic fields are generated in our model preferably on large scales, whereas small scale fields are strongly suppressed. This result is true at least in the non-saturated regime and we are currently exploring the saturated case. Such a property is highly interesting again with respect to dynamo amplification theories: the small scale fields are weak and therefore their amplification by dynamo represent a less crucial threat.

We showed also that, in the context of the ISM, our mechanism is negligible as a source of magnetic fields when compared to usual effects such as dynamo and turbulence.

As we mentioned earlier, Subramanian \textit{et. al.}~\cite{subra} and Gnedin \textit{et. al.}~\cite{gnedin} explored the generation of magnetic field seeds in the context of reionisation. In their model, the magnetic field source term is the Battery effect relying on charge separation. The latter is obtained when ionisation fronts either break out from protogalaxies or propagate past overdense regions of the Universe. Across an ionising front, the temperature varies sharply by several orders of magnitude. The driving physical process for charge separation is the thermal pressure of the gas, acting in the same way on electrons and ions. Since their mass is far smaller than the ion mass, electrons are more accelerated by the pressure force than ions and this provides charge separation. However, the mechanism we presented in this paper relies on a quite different process: The source of charge separation is not differential thermal pressure but differential radiation pressure. For a particle $i=e,p$, the acceleration due to thermal pressure is $a^\text{t}_i \propto m_i^{-1}$ whereas radiation pressure provides an acceleration $a^\text{r}_i \propto m_i^{-3}$. Radiation pressure is therefore more efficient by a factor of 
\beq
\frac{a^\text{r}_\text{e}/a^\text{r}_\text{p}}{a^\text{t}_\text{e}/a^\text{t}_\text{p}} = \left(\frac{m_\text{p}}{m_\text{e}}\right)^2 \approx 3. ~10^6
\eeq
 than thermal pressure for accelerating electrons and creating electrical fields. This explains why our model leads to orders of magnitude higher predictions than in \cite{subra}. Thus, the mechanism we described in this paper is likely to be dominant in cosmological situations where strong anisotropic and inhomogeneous radiation pressure exists. 

Further developments of the model presented in this paper,  relaxing our geometrical assumptions and addressing the magnetic field power spectrum  in the saturated regime will appear in a forthcoming paper.\\

\begin{acknowledgments}
We thank Dr. M. Tagger for having drawn our attention to the potential importance of the saturated regime. We also thank Dr. S. Colombi for comments, and an anonymous referee for useful remarks and interesting references.
\end{acknowledgments}

\bibliography{magnet}
\end{document}